\documentclass[a4paper,twocolumn,notitlepage,aps,pra,10pt]{revtex4-2}
\usepackage{bbm}
\usepackage{amsmath}
\usepackage{amssymb}
\usepackage{amsthm}
\usepackage{amsfonts}
\usepackage{amsmath}
\usepackage{graphicx}
\usepackage{xcolor}
\usepackage{csquotes}
\usepackage{braket}
\usepackage{setspace}
\usepackage{verbatim}
\usepackage[colorlinks=true,linkcolor=teal,citecolor=teal,urlcolor=teal]{hyperref}
\usepackage{mathtools}
\usepackage{booktabs}
\usepackage{braket}
\usepackage{mleftright}
\usepackage{siunitx}
\usepackage[normalem]{ulem}
\usepackage[ruled,lined]{algorithm2e}
\usepackage{tabularx}
\SetKw{Continue}{continue}

\DeclareSIUnit\hartree{Ha}

\theoremstyle{definition}

\newcommand{\ba}{\begin{eqnarray}}
\newcommand{\ea}{\end{eqnarray}}
\newcommand{\be}{\begin{equation}}
\newcommand{\ee}{\end{equation}}
\newcommand{\beq}{\begin{equation}}
\newcommand{\eeq}{\end{equation}}
\newcommand{\bba}{\begin{eqBox}\begin{eqnarray}}
\newcommand{\eba}{\end{eqnarray}\end{eqBox}}

\newcommand{\ie}{{\it{i.e.}~}}

\newcommand{\Tr}[1]{{\rm Tr} \left[ {#1} \right]}
\newcommand{\overbar}[1]{\mkern 1.5mu\overline{\mkern-1.5mu #1 \mkern-1.5mu}\mkern 1.5mu}

\begin{document}

\title{Enhanced observable estimation through classical optimization of informationally over-complete measurement data --- beyond classical shadows}

\author{Joonas Malmi}
\email{joonas@algorithmiq.fi}
\author{Keijo Korhonen}
\author{Daniel Cavalcanti}
\author{Guillermo Garc\'{\i}a-P\'{e}rez}
\affiliation{Algorithmiq Ltd, Kanavakatu 3 C, FI-00160 Helsinki, Finland}

\date{\today}

\begin{abstract}
In recent years, informationally complete measurements have attracted considerable attention, especially in the context of classical shadows. In the particular case of informationally over-complete measurements, for which the number of possible outcomes exceeds the dimension of the space of linear operators in Hilbert space, the dual POVM operators used to interpret the measurement outcomes are not uniquely defined. In this work, we propose a method to optimize the dual operators after the measurements have been carried out in order to produce sharper, unbiased estimations of observables of interest. We discuss how this procedure can produce zero-variance estimations in cases where the classical shadows formalism, which relies on so-called canonical duals, incurs exponentially large measurement overheads. We also analyze the algorithm in the context of quantum simulation with randomized Pauli measurements, and show that it can significantly reduce statistical errors with respect to canonical duals on multiple observable estimations.
\end{abstract}

\maketitle

The study of quantum physics and the development of quantum technologies are based on our ability to extract useful information from quantum systems.
In particular, quantum simulation on quantum computers typically requires performing measurements from which we can infer physical properties such as energy, magnetization, entropy or correlations.
A standard procedure is to perform quantum state tomography and obtain a description of the quantum state, from which we can estimate any observable or quantum-information-theoretical quantity.
However, the number of parameters needed for such a task generally grows exponentially with the number of constituents of the system.
This implies that the measurement cost (either in terms of measurement settings or shots) required to reach a certain precision, and the classical memory to store the data, become unattainable even for small system sizes. 

An efficient way to estimate the mean value of different observables without the need to reconstruct the full quantum state is to apply an informationally complete measurement, given by informationally complete positive operator-valued measures (IC-POVMs), and classically post-process the data using the dual effects of the measurement~\cite{D_Ariano_2007, general-measurement-frames}.
This idea recently attracted significant attention after the realization that particular choices of IC-POVMs and dual effects (called \emph{classical snapshots} in the framework of \emph{shadow estimation}~\cite{classical-shadows}) can lead to efficient estimations in the number of measurement shots and qubits~\cite{classical-shadows}.
Several works have proposed other classes of IC-POVMs \cite{importance-sampled-shadow,2020arXiv200615788H, Shadow-depth,Shallow-shadows, Brickwork-shadow,Shadow-Guhne}, and even on-the-fly optimization procedures \cite{learning-to-measure,Shadow-Guhne,Adam}.

While previous works on this type of estimation focused on proposing different IC-POVMs (which correspond to different measurement setups), refs.~\cite{D_Ariano_2007, state-overlap, general-measurement-frames} recognized that the accuracy of the estimations depends on the dual POVM effects used in the estimator as well.
Here we make this idea practical and introduce a method to optimize the dual effects of a POVM in a practical scenario where one has access to a finite sample of measurement data.
We show that this method can provide exponential advantages in measurement overhead with respect to local shadow estimation, and apply it to a variety of physically relevant problems ranging from spin chain dynamics to quantum chemistry calculations.

Let us first review the idea of observable estimation through informationally complete POVMs.
A POVM is described by positive operators (also called POVM effects) $\Pi_i\geq 0$ ($i=0,...,r-1$) that add up to identity, \ie $\sum_{i=0}^{r-1} \Pi_i = \mathbb{I}$.
Upon measuring a state $\rho$ with a POVM we obtain an outcome $i$ with probability $p_i=\Tr{\Pi_i \rho}$, so that we have $r$ possible results, or outcomes.
A particularly important set of POVMs is informationally complete POVMs (IC-POVMs), for which the POVM effects span the space of linear operators in the Hilbert space, $\mathcal{L} (\mathcal{H})$.
This means that an IC-POVM needs to have $r\geq d^2$ effects, where $d^2$ effects are linearly independent.
Thus, we can write any operator $O\in \mathcal{L} (\mathcal{H})$ as $O=\sum_i c_i \Pi_i$.
If an IC-POVM is IC and has exactly $r=d^2$ linearly independent effects, it is called a minimal IC-POVM.

As mentioned above, IC-POVMs can be used to estimate the mean value of different observables while bypassing the explicit reconstruction of the quantum state \cite{D_Ariano_2007}.
This is done by first noticing that every IC-POVM can be associated with a set of \emph{dual effects} $D_i$ ($i=0,...,r-1$) that are defined by operators satisfying 
\begin{equation}\label{eq-dual-definition}
O=\sum_i \Tr{O\Pi_i} D_i
\end{equation}
for every operator $O$.
This means that $\{D_i\}_{i=1}^{r}$ also form the dual operator basis in $\mathcal{L} (\mathcal{H})$.
(Consequently, they are sometimes called \emph{dual frames}.)
If we choose $O$ to be a quantum state $\rho$ we obtain $\rho=\sum_i p_i D_i$.
This means that $\rho$ is the averaged dual POVM, where the average is taken with respect to the measurement outcome probabilities $\{ p_i = \Tr{\rho \Pi_i}\}$.
Thus, in a real experiment where we have access to $S$ experimental shots of an IC-POVM, we can construct an unbiased estimator of $\rho$ as $\overbar{\rho}=\sum_i f_i D_i$, where $f_i$ is the observed frequency of outcome $i$~\footnote{To see that $\overbar{\rho}$ is unbiased, $\mathbb{E} [\overbar{\rho}] = \rho$, notice that $\mathbb{E} [\overbar{\rho}] = \sum_i \mathbb{E} [f_i] D_i = \sum_i p_i D_i$, since $f_i = s_i / S$, where $s_i$ is the number of times outcome $i$ was obtained out of the $S$ experiments. The random variables $s_i$ follow a mutinomial distribution with mean $Sp_i$.}.
Furthermore, we can also define unbiased estimators of different observables as  
\begin{equation}\label{eq-mean-estimator}
\overbar{O}=\Tr{O \overbar{\rho}}=\sum_i f_i \Tr{O D_i}.
\end{equation} 
These estimators are also consistent, that is, they converge to $\langle O \rangle$ in the limit of $S\rightarrow \infty$ since $f_i\rightarrow p_i$.

As an illustrative example, let us consider the shadow estimation of qubit observables proposed in ref.~\cite{classical-shadows}.
Their proposed measurement scheme consists in measuring the qubit in the $Z, X$, or $Y$ Pauli basis with equal probability.
The POVM effects are given by:
\begin{align}\label{eq-random-pauli}
    &\Pi_0 = \frac{1}{3} |0 \rangle \langle 0 |,\quad \Pi_1 = \frac{1}{3} |1 \rangle \langle 1 |, \quad \Pi_2 = \frac{1}{3} | + \rangle \langle + | \\
    &\Pi_3 = \frac{1}{3} | - \rangle \langle - |,\quad \Pi_4 = \frac{1}{3} | +i \rangle \langle +i |,\quad \Pi_5 = \frac{1}{3} | -i \rangle \langle -i |,\nonumber
\end{align}
where $| \pm \rangle = 1 / \sqrt{2} (|0\rangle \pm | 1\rangle)$ and $| \pm i \rangle = 1 / \sqrt{2} (|0\rangle \pm i|1\rangle)$.
The particular set of duals considered in ref.~\cite{classical-shadows}, dubbed \emph{classical shadows} therein, are
\begin{align}\label{eq-canonical-duals}
    &D_0 = \frac{1}{2} ( I + 3 Z ),\quad D_1 = \frac{1}{2} ( I - 3 Z ),\quad D_2 = \frac{1}{2} ( I + 3X ), \nonumber \\
    &D_3 = \frac{1}{2} ( I - 3X ),\quad D_4 = \frac{1}{2} (I + 3 Y ),\quad D_5 = \frac{1}{2} (I - 3 Y ).
\end{align}
We will refer to these duals as \textit{canonical duals} hereafter.

Notice that the POVM \eqref{eq-random-pauli} is IC, but it is composed of more than $d^2$ effects, so it is over-complete (OC-POVMs for short).
Consequently, some of the effects are linearly dependent on the others.
As we will show next, the duals of OC-POVMs are not uniquely determined, and this freedom of choice can be used to improve the estimation.

To find the duals of an $r$-outcome OC-POVM, let us first choose $d^2$ linearly independent effects among the ones of the POVM.
We will call them the \emph{basis effects}, and denote them by $\{\overbar{\Pi}_{i}\}_{i=0}^{d^2-1}$.
The remaining $r-d^2$ effects will be called the \emph{redundant effects}, and will be denoted by $\{\widetilde{\Pi}_{i}\}_{i=d^2}^{r-1}$.
Similarly, to these effects we will associate the \emph{basis dual effects} $\{\overbar{D}_{i}\}_{i=0}^{d^2-1}$ and the \emph{redundant dual effects} $\{\widetilde{D}_{i}\}_{i=d^2}^{r-1}$. 
Using this notation, Eq.~\eqref{eq-dual-definition} reads
\begin{equation}
   O = \sum_{i=0} ^{d^2 -1} \Tr{O \overbar{D}_i} \overbar{\Pi}_i + \sum_{j=d^2} ^{r-1} \Tr{O \widetilde{D}_j} \widetilde{\Pi}_j.
\end{equation}

Notice that the basis effects form an (unnormalized) minimal IC-POVM, so we can write $\widetilde{\Pi}_{j} = \sum_{i=0} ^{d^2 - 1} \Tr{D_i^\star \widetilde{\Pi}_j} \overbar{\Pi}_i$, where we have used the symbol $\star$ to denote the unique duals to said minimal basis.
Thus, we get
\begin{equation}\label{eq-linear-decomp-rewrite}
\begin{split}
   O & = \sum_{i=0} ^{d^2 - 1}\Big( \Tr{O \overbar{D}_i} + \sum_{j=d^2} ^{r-1} \Tr{O \widetilde{D}_j} \Tr{D_i^\star \widetilde{\Pi}_j} \Big) \overbar{\Pi}_i \\
    & = \sum_{i=0} ^{d^2 -1} \Tr{O {D}_i^\star} \overbar{\Pi}_i.
\end{split}
\end{equation}
The last term in the expression is the unique decomposition of $O$ in terms of the basis dual effects.
Since this equality must hold for any operator $O$, the dual effects must fulfill
\begin{equation}\label{eq:constraint}
    \overbar{D}_i + \sum_{j=d^2} ^{r-1} \widetilde{D}_j \Tr{D_i^\star \widetilde{\Pi}_j} = {D}_i^\star.
\end{equation}
This constraint between basis and redundant dual effects can be automatically satisfied by writing the former in terms of the latter, so we can parameterize the full set of duals of an OC-POVM as
\begin{equation}\label{eq-full-duals}
   \begin{cases}
        D_i^\star - \sum_{j=d^2} ^{r-1} \Tr{D_i^\star \widetilde{\Pi}_j} \widetilde{D}_j,\ i=0,\cdots,d^2-1 \\
        \widetilde{D}_j,\ j=d^2,\cdots,r-1,
    \end{cases}
\end{equation}
where $\widetilde{D}_j$ are Hermitian matrices that can be chosen freely.

We may exploit this freedom of choice to find the dual effects in such a way that the \emph{variance} of the estimator of an observable $O$ is minimized.
The per-shot variance when estimating $O$ from the POVM outcomes is given by
\begin{equation}\label{eq:variance}
    {\rm Var}[O] = \sum_i p_i \big(\Tr{O D_i}\big)^2 - \Big(\sum_i p_i  \Tr{O D_i} \Big)^2,
\end{equation}
where $p_i$ is the probability of obtaining the $i$-th outcome.
While the second term does not depend on the choice of dual effects, as long as these satisfy Eq.~\eqref{eq:constraint}, the first term---the second moment of $\Tr{O D_i}$ with respect to the probability distribution of the outcomes---does.
Therefore, by choosing dual effects that minimize the second moment, the statistical errors in the estimation of the expectation value of the operator $O$ may be significantly smaller than with e.g.~the canonical duals, even when using the same measurement outcome data.

As a simple example of how optimizing the duals of OC-POVMs can provide an advantage, consider the task of estimating the average value of the Pauli-Z observable for the state $| 0 \rangle$ using the measurement in Eq.~\eqref{eq-random-pauli}.
One can readily see that the probabilities to obtain the different outcomes are $p_0 = 1/3$, $p_1 = 0$, $p_2 = p_3 = p_4 = p_5 = 1/6$.
Using the canonical duals in Eq.~\eqref{eq-canonical-duals}, we obtain the variance ${\rm Var}[Z] = \sum_i p_i (\Tr{D_i Z}) ^2 - \Big( \sum_i p_i \Tr{D_i Z} \Big)^2 = 2$.
Instead, if we optimize the duals to minimize the variance, we find ${\rm Var}[Z]=0$ with the choice of duals
\begin{align*}
    &D_0 ^{\text{opt}} = \frac{1}{2} (I + Z),\quad D_1 ^{\text{opt}} = \frac{1}{2} (I - 5 Z), \\
    &D_2 ^{\text{opt}} = \frac{1}{2} (I + 3 X +  Z),\quad  D_3 ^{\text{opt}} = \frac{1}{2} (I - 3X + Z ), \\
    &D_4 ^{\text{opt}} = \frac{1}{2} (I + 3 Y + Z), \quad  D_5 ^{\text{opt}} = \frac{1}{2} (I - 3Y + Z ).
\end{align*}
Notice that in the case of a multi-qubit state $| 0\rangle ^{\otimes N}$ and an observable $Z ^{\otimes N}$, the canonical duals \eqref{eq-canonical-duals} in ref.~\cite{classical-shadows} result in an exponential overhead, with variance ${\rm Var}[Z]=3^N - 1$, while the optimal duals result in zero-variance estimations.

In this particular example, since the OC-POVM is implemented by measuring in the Pauli basis and the operator is a Pauli operator, one could alternatively achieve a zero-variance estimation by considering the measurement outcomes in which the qubits are measured in the $Z$ basis, discarding all other data.
However, dual optimization is a general-purpose approach that can be used with more complex OC-POVMs as well, for which no simple and efficient data post-processing strategy may be obvious.
This is particularly relevant in schemes in which one optimizes overcomplete measurement setups, as in ref.~\cite{Adam}.

While the above discussion introduces the basic idea of dual optimization and its potential, its practical implementation poses additional challenges.
In general, the probability distribution of the outcomes $\{p_i\}$ in the second moment $\sum_i p_i \big(\Tr{O D_i}\big)^2$ is not knowable, since its characterization is exponentially hard.
In practice, after repeating the measurement a finite number of times, we can only estimate it based on the observed frequencies
\begin{equation}\label{eq:second-moment-est}
    \overbar{\Tr{O D_i}^2} = \sum_i f_i \big(\Tr{O D_i}\big)^2.
\end{equation}
We propose to minimize this quantity.
Since in the finite statistics scenario the first moment $\sum_i f_i \Tr{O D_i}$ depends on the duals used, the estimation with the duals the minimize Eq.~\eqref{eq:second-moment-est} should typically have smaller statistical error.
However, doing so may introduce statistical biases: if the dual effects $\{ D_i \}$ are modified as to minimize $\sum_i f_i \big(\Tr{O D_i}\big)^2$, then the duals and the measurement data $f_i$ are no longer statistically independent, which means that we cannot guarantee that the estimator of the mean $\sum_i f_i \Tr{O D_i}$ is unbiased.
In order to prevent this, our procedure consists in splitting the measurement data into two disjoint sets, $A$ and $B$.
We first optimize the duals using a \emph{training dataset} $A$, and then evaluate the mean and the variance using the optimized duals in an \emph{estimation dataset} $B$.
Since $A$ and $B$ are disjoint, the optimized duals and the frequencies used in the final estimation are statistically independent.
We then repeat the procedure swapping the roles of $A$ and $B$, hence producing another pair of estimations of the mean and the variance that are then combined with the former ones, so no data is left unused.
The optimization of the duals is performed by sweeping over the qubits, optimizing the duals of each qubit one at a time, as we observed this strategy to converge faster than a global optimization approach.
In what follows, each single-qubit optimization is carried out using the L-BFGS optimizer.

In order to test the approach, we consider two physically motivated problems: the estimation of the real part of 2-reduced density matrix (2-RDM) operator elements of the ${\rm H_2O}$ and ${\rm LiH}$ molecules, and the quantum simulation of a spin chain performed in ref.~\cite{PEC}.
In what follows, we compare the absolute error of the estimations, defined as $\epsilon = |\langle O \rangle - \overline{O}|$, with and without dual optimization.

\begin{figure}
    \centering
    \includegraphics[width=
    0.8 \columnwidth]{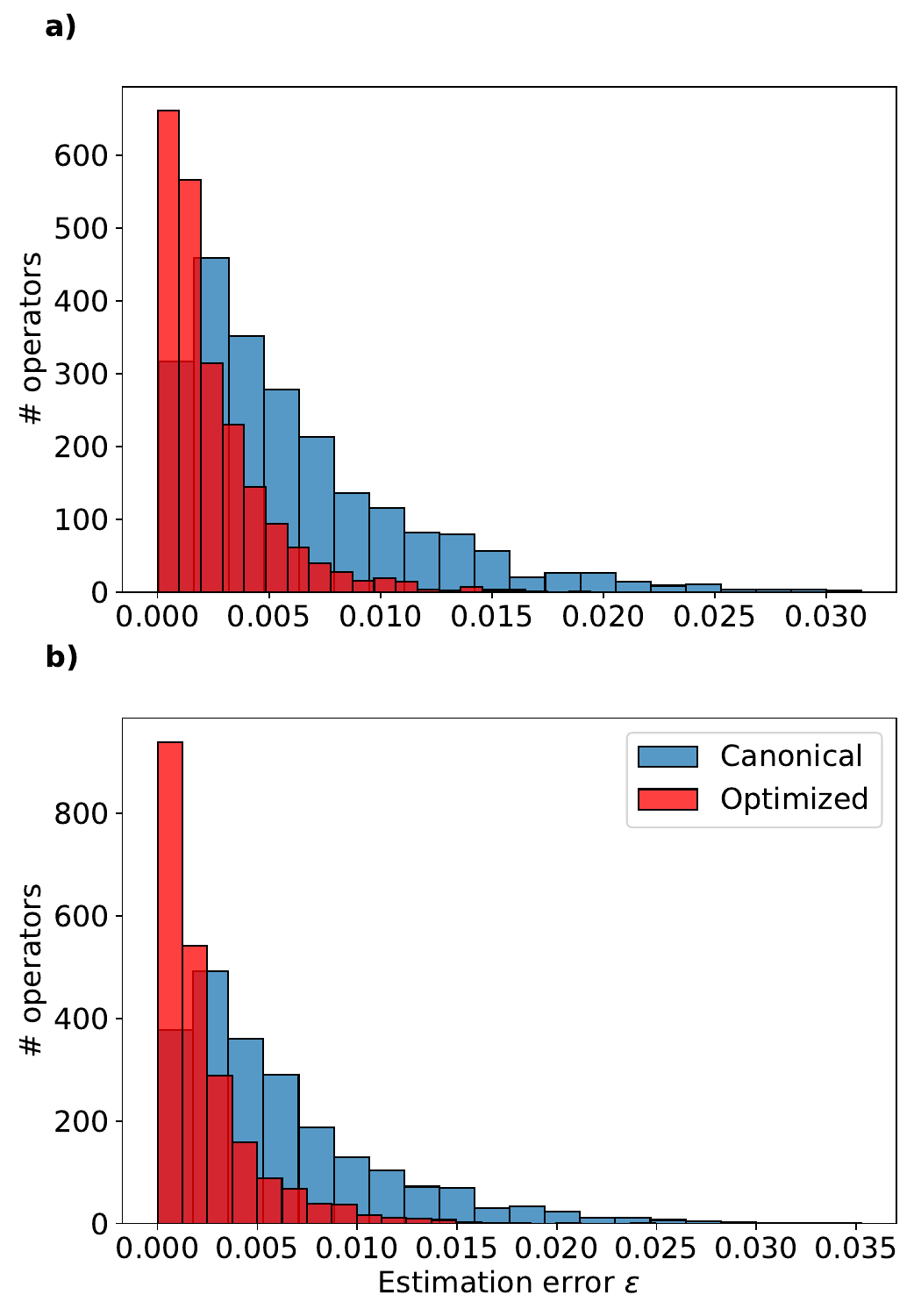}
    \caption{Distribution of absolute errors $\epsilon$ over the real parts of fermionic 2-RDM elements for the ${\rm H_2O}$ (a) and ${\rm LiH}$ (b) molecules.
    Pre-trained VQE states are measured using the POVM in Eq.~\eqref{eq-random-pauli}.
    The resulting $2 \times 10^6$ shot dataset is then used to evaluate all the 2-RDM expectation values with the canonical duals, as well as to optimize the duals using the protocol introduced in the main text and produce the corresponding estimations.}
    \label{fig:rdms}
\end{figure}

We first consider the problem of estimating the real part of fermionic 2-reduced density matrix (2-RDM) operator elements $a_i^\dagger a_j^\dagger a_k a_l$, where $a_i$ are fermionic annihilation operators.
To that end, we consider the ${\rm H_2O}$ and ${\rm LiH}$ molecules in a minimal basis set.
Their second quantized Hamiltonians and 2-RDM elements are mapped to qubit space using the JKMN fermion-to-qubit mapping \cite{Jiang2020optimalfermionto}, resulting in 12-qubit operators.
We then use pre-trained hardware-efficient variational quantum eigensolver (VQE) states~\cite{hardware-efficient-ansatz-HEA} to approximate their ground states, and measure these states with the randomized Pauli measurements in Eq.~\eqref{eq-random-pauli} to obtain $2 \times 10^6$ shots.
The resulting dataset is then used to evaluate all the 2-RDM elements with canonical duals, as well as to optimize them and produce the corresponding estimations.
As it can be seen in Fig.~\ref{fig:rdms}, where we depict the distribution of $\epsilon$ over RDM elements, the optimization has the overall effect of significantly reducing statistical errors.

\begin{figure}
    \centering
    \includegraphics[width=0.8
    \columnwidth]{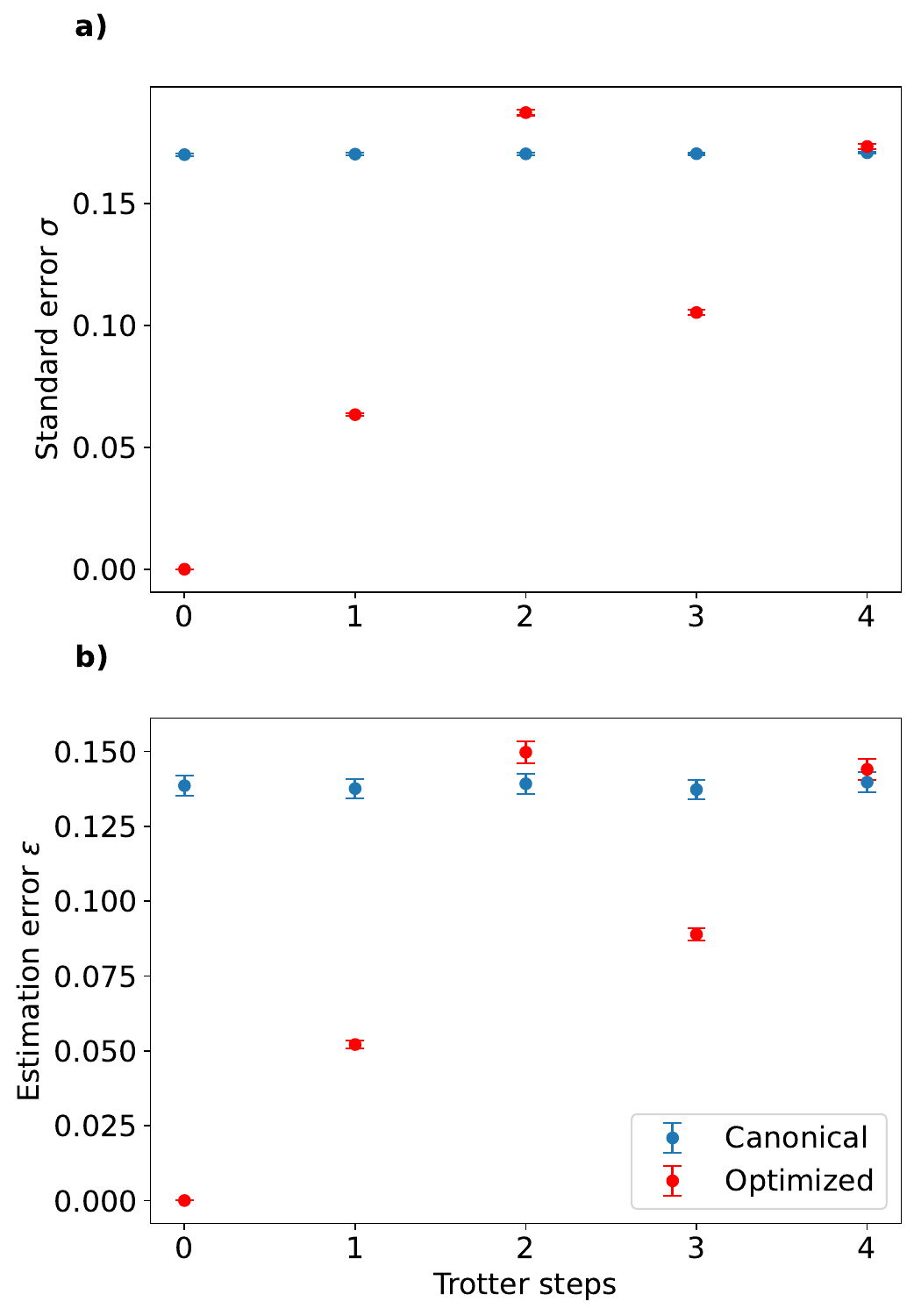}
    \caption{Estimated standard error $\sigma = [(\sum_i f_i \Tr{O D_i}^2 - \overline{O}^2)/N_{\rm shots}]^{1/2}$, where $N_{\rm shots}$ is the number of shots used in the estimation, and absolute error (a) and b), respectively) of the observable $O = Z ^{\otimes 10}$ for different Trotter steps in a 10-qubit system.
    At each time step, the state is measured 1000 times using 2 million measurement shots per repetition.
    In each repetition, the data set is split in two one-million-shot datasets A and B.
    First, dataset A is used for optimization and dataset B for estimation, and then vice versa.
    The values shown here are averaged over the estimation values between the two data sets.
    For these results, we performed 20 sweeps over the qubits during the optimization.
    The average and standard deviation over repetitions.}
    \label{fig:trotter}
\end{figure}

As a second application, in Fig.~\ref{fig:trotter}, we show the results for the simulation of the Trotter evolution of a 10-spin transverse-field Ising model performed in ref.~\cite{PEC}. The Hamiltonian reads
\begin{equation}
    H=-J \sum_i Z_i \otimes Z_{i+1} + h \sum_i X_i,
\end{equation}
with $J=0.5236$ and $h=1$, with initial state $|0 \rangle^{\otimes 10}$.
The estimated observable is $O = Z ^{\otimes 10}$.
For each Trotter step, we repeat the simulation 1000 times, sampling $2 \times 10^6$ shots per simulation.

As we can see, dual optimization provides a reduced error for steps 0, 1 and 3.
In particular, at step 0, the procedure consistently achieves a zero-variance estimation.
This is possible because the state and observable are $\ket{0}^{\otimes N}$
and $Z^{\otimes N}$, respectively, which is precisely the aforementioned example of an exponential overhead with canonical duals.
However, notice that the algorithm consistently finds optimal duals based only on the provided measurement data, assuming the canonical duals as a starting point.
For steps 2 and 4, the optimization procedure leads to higher $\sigma$ and $\epsilon$.
This is a direct consequence of over-fitting: while the optimized duals reduce the second moment Eq.~\eqref{eq:second-moment-est} for the training dataset, they increase it for the estimation dataset.
In order to mitigate this effect, one can simply monitor the value for the estimation set and stop the optimization as soon as the training and estimation dataset values differ significantly.
In any case, notice that the estimated standard error $\sigma$ is an observable quantity that does not require previous knowledge of $\langle O \rangle$ (as opposed to the absolute error $\epsilon$) and, as Fig.~\ref{fig:trotter} shows, $\sigma$ is a good proxy of the actual estimation error $\epsilon$.
Therefore, one can always use the estimation provided by the duals with smallest standard error on the estimation dataset.

In this paper, we have shown that the estimation of physical properties by means of informationally over-complete POVMs can be greatly improved by the optimization of the dual effects used in the estimator.
This improvement comes in the form of a reduced estimation variance, which in turn provides an advantage in terms of measurement overhead (that is, the number of shots required to achieve a certain precision).
Furthermore, this optimization is purely classical, and can be performed in post-processing, thus not requiring any modification of the physical setup.

We have demonstrated the feasibility of our optimization procedure in physically relevant problems, such as the Trotter evolution of a spin system, and different physical properties of molecular systems.
We believe that our method can be of great help not only for the estimation of physical properties, but also for subroutines of quantum computing and simulation protocols.
For instance, in some VQE approaches, one needs to estimate a great number of commutators~\cite{grimsley2019adaptive}, which typically results in a prohibitive measurement cost.
Another example is the estimation of stabilizer observables for error correction codes and other applications such as one-way quantum computation.
This method can also be used in conjunction with adaptive POVMs, where the POVM is first optimized with respect to some state and operators, and the duals are then optimized in post-processing as well. 

Finally, the numerical simulations presented here only considered the POVMs given in \ref{eq-random-pauli} for simplicity. However, the methodology is generally valid for any $N$-qubit product OC-POVM, so it will be interesting to explore to what extent dual optimization can improve multi-observable estimation for other POVM classes (for instance, with more outcomes).

\begin{acknowledgements}
We thank Sergey Filippov and Boris Sokolov for helpful discussions.
While finishing this project, we became aware of related work by L. Fischer \textit{et.~al}, in which they also observe estimation improvements using dual optimization for randomly chosen states and observables.
\end{acknowledgements}

\bibliography{bibliography}

\end{document}